\documentclass{article}
\usepackage{amssymb,amsmath,cite,bm,graphicx,xfrac}

\def\be{\begin{equation}}
\def\ee{\end{equation}}
\def\bea{\begin{eqnarray}}
\def\eea{\end{eqnarray}}

\begin{document}
\begin{titlepage}
\begin{center}
{\large \textbf{
Regge Trajectories by 0-Brane Matrix Dynamics}}

\vspace*{2\baselineskip}

Amir~H.~Fatollahi
\\
\vspace{\baselineskip}
\textit{ Department of Physics, Alzahra University, \\
Vanak 1993893973, Tehran, Iran}\\
\vspace{\baselineskip}
\texttt{fath@alzahra.ac.ir}
\end{center}
\vspace{\baselineskip}
\begin{abstract}
\noindent The energy spectrum of two 0-branes for fixed angular momentum
in 2+1 dimensions is calculated by the Rayleigh-Ritz method.
The basis function used for each angular momentum consists of 80 eigenstates
of the harmonic oscillator problem on the corresponding space. It is seen that the
spectrum exhibits a definite linear Regge trajectory behavior.
It is argued how this behavior supports 
the picture by which the bound-states of quarks and QCD-strings are governed
by the quantum mechanics of matrix coordinates.
\end{abstract}

\vspace{2\baselineskip}


\end{titlepage}

\section{Introduction}
The string theoretic description of gauge theories is an old idea
\cite{Polya,wilson,largen}, still stimulating research works in theoretical physics
\cite{Polyakov, adscft,Po3}.
Depending on the amount of momentum transfer, the hadron-hadron scattering
processes have shown two different behaviors \cite{Close,roberts}.
At very large momentum transfers the interactions are among the
point-like substructures, and qualitative similarities to
electron-hadron scattering emerge.  At high energies
and small momentum transfers the Regge trajectories are exchanged. The exchanged
linear trajectories are the first motivation for the string picture of
strong interaction. However, the fairly good fitting between the linear
Regge trajectories and the mass of QCD bound-states has not been explained yet \cite{Po3},
partially due to the lack of a consistent formulation of string theory in 3+1 dimensions

According to string theory, 0-branes are point-like objects to which the
strings can end \cite{Po1,Po2}. It is known that in a specific regime the dynamics of
$N$ 0-branes is governed by the matrix quantum mechanics resulting from
dimensional reduction of U$(N)$ Yang-Mills theory to $0+1$ dimension
\cite{9510135}. In this regime, the dynamics of 0-branes and
the strings stretched between them is encoded in the elements of matrix
coordinates resulted from the dimensional reduction
of non-Ableian gauge theory.

By the picture mentioned above, it sounds reasonable that the dynamics
of 0-branes is used to model the bound-states of quarks
and QCD-strings. This picture is the main theme of a series of works, and
it is shown that the dynamics of 0-branes can reproduce some
known features and expectations in hadron physics, including the potentials
between static and fast decaying quarks, and also the Regge behavior in 
the scattering amplitudes \cite{fat1,fat2}.
The symmetry aspects of the picture were studied in \cite{fat3}.
In particular, it is argued that maybe the full featured formulation of non-Ableian
gauge theories is possible on non-commutative matrix spaces \cite{fat3}.

In the present note the aim is to see whether the 0-brane matrix dynamics can
reproduce the linear Regge trajectories observed in hadron physics.
Early studies on spectrum of 0-brane bound-states are reported in \cite{halpern,nicolai,danielsson,kabat}.
In \cite{wosiek,kares} the study of spectrum based on the variational method is presented.
In the present work, based on the results by \cite{kares,kabat},
for the case of two bosonic 0-branes in 2+1 dimensions the energy eigenvalues
are calculated for states with given angular momentum, ranging from 0 to 42. The
spectrum is calculated by the Rayleigh-Ritz variational method, and the basis function
for each angular momentum consists of 80 eigenstates of the harmonic oscillator
problem on the configuration space of the 0-branes.
It is seen that apart from two lowest angular momenta, the energy versus
angular momentum can be fitted with straight-line at each level.
The spectrum may be interpreted as the one for massive 0-branes in 2+1 dimensions, or
in a Matrix theory perspective \cite{9610043}, as for massless
particles in 3+1 dimensions but in the light-cone frame.
Based on the latter way of interpretation, the linear
relation can be turned as the one between mass-squared and angular momentum,
just reminiscent the observed one in QCD bound-states.

Based on the above observation about the spectrum, this may be suggested that,
the quantum mechanics of matrix coordinates
can reconcile string picture and QCD in 3+1 dimensions. In particular,
according to this picture the bound-states of quarks and QCD-strings are governed
by the quantum mechanics of matrix coordinates \cite{fat1,fat2,fat3}.

The scheme of the rest of this paper is the following. In Sec.~2,
the basic notions for the 0-brane matrix dynamics are presented,
together with a demonstration of a bound-state classical solution.
In Sec.~3, the quantum theory is developed. The eigen-functions
of the angular momentum together with the complete solution for
harmonic oscillator on the 0-branes' configuration space are presented.
This solution is used to construct the basis function used in the
Rayleigh-Ritz variational method of Sec.~4.
The light-cone interpretation of the results is also presented in Sec.~4.

\section{Matrix Dynamics of 0-Branes}

The dynamics of $N$ 0-branes is given by a U$(N)$ Yang-Mills theory dimensionally
reduced to $0+1$ dimensions \cite{Po2,kabat}, given by (in units $\hbar=c=1$)
\bea\label{1}
&~&L=m_0 \mathrm{Tr}\; \biggl(\frac{1}{2}  (D_tX_i)^2 + \frac{1}{4\,l_s^4}\,[X_i,X_j]^2\biggl),\\
&~&i,j=1,...,d,\;\;\;\;\;\; D_t=\partial_t-\mathrm{i}[A_0,\;],\nonumber
\eea
with $l_s$ as the fundamental string length, and $m_0=(g_sl_s)^{-1}$, with
$g_s$ as the supposedly small string coupling, \textit{i.e.} $m_0\gg l_s^{-1}$.
$X$'s are in adjoint representation of U$(N)$ with the usual expansion
$X_i=x_{i\, a} T_{\, a}$, $\, a=1,..., N^2$.
The theory is invariant under the gauge symmetry
\bea\label{2}
\vec{X}&\rightarrow&\vec{X'}=U\vec{X}U^\dagger,\nonumber\\
A_0&\rightarrow&A'_0=UA_0U^\dagger+\mathrm{i} U\partial_tU^\dagger,
\eea
where $U$ is an arbitrary time-dependent $N\times N$ unitary matrix. Under
these transformations one can check that:
\bea\label{3}
D_t\vec{X}&\rightarrow&D'_t\vec{X'}=U(D_t\vec{X})U^\dagger,\nonumber\\
D_tD_t\vec{X}&\rightarrow&D'_tD'_t\vec{X'}=U(D_tD_t\vec{X})U^\dagger.
\eea
For each direction there are $N^2$ variables and it is understood that
the extra $N^2-N$ degrees of freedom are representing the dynamics
of oriented strings stretched between $N$ 0-branes. The center-of-mass of 0-branes
is represented by the trace of the $X$ matrices.

In the quantum theory the off-diagonal elements of matrices play an essential role. In particular,
it is shown that in the quantum theory the off-diagonal elements cause the interaction
between 0-branes. For the case of classically static 0-branes it is shown that the
fluctuations of the off-diagonal elements develop a linear potential,  just as the
case for QCD-strings stretched between quarks \cite{fat1}.

The canonical momenta are given by:
\bea\label{4}
P_{i}=\frac{\partial L}{\partial X_{i}}=m_0\,D_t {X}_i
\eea
by which the Hamiltonian is constructed
\bea\label{5}
H=\mathrm{Tr}\;\biggl(\frac{P_i^2}{2\, m_0}- \frac{m_0}{4\,l_s^4}\,[X_i,X_j]^2\biggl).
\eea
As the time-derivative of the dynamical variable $A_0$ is absent, its
equation of motion introduces a constraint, the so-called Gauss's law
\bea\label{6}
G_a:=\sum_i [X_i,P_i]_a=\mathrm{i}\sum_{i,b,c}  f_{abc}\,x_{i\,b}\,p_{i\, c}\equiv 0.
\eea
In the present work we take the two dimensional case ($d=2$) for a pair of 0-branes.
It would be quite useful to separate the pure gauge variables from the others.
For the case of SU(2) theory in 2+1 dimensions,
following \cite{kares,halpern} we use the decomposition
\begin{align}\label{7}
x_{i\, a}=(\Psi)_{a\,b}(\Lambda)_{b\,j}(\eta)_{j\,i}
\end{align}
in which the matrix $\Psi$ is an element of group of SU(2).
Accordingly the gauge transformations of the variable $x_{i\,a}$ are captured by $\Psi$
through ordinary gauge group left multiplications. Parameterizing the SU(2)
group elements by the three Euler angles, the matrix $\Psi$ is represented by \cite{goldstein}
\begin{align}\label{8}
\Psi=R_z(\alpha)R_x(\gamma)R_z(\beta),
\end{align}
in which $R_a$ is the rotation matrix about the $a\,\!$th axis. Analogously, the matrix
$\eta$ is an element of the SO(2) group parameterized by the angle $\phi$,
capturing the effect of rotation
in the two dimensional space. The matrix $\Lambda$ takes the form \cite{kares}
\begin{align}\label{9}
\Lambda=\begin{pmatrix}r\cos\theta & 0 \cr
0 & r\sin\theta\cr
0 & 0
\end{pmatrix}
\end{align}
We mention that the only variable with dimension of length is $r$. Also, apart from pure gauge
variables $\alpha$, $\beta$, and $\gamma$, the two dimensional configuration space is spanned by the polar coordinates $(r,\phi)$,
and the extra variable $\theta$ appears as an internal degree of freedom.

By the decomposition, the three constraints (\ref{6}) take the form \cite{kares}:
\begin{align}\label{10}
G_1&=\sin\alpha\cot\gamma~ p_\alpha-\sin\alpha\csc\gamma~ p_\beta -\cos\alpha~ p_\gamma\cr
G_2&=\cos\alpha\cot\gamma~ p_\alpha -\cos\alpha\csc\gamma ~p_\beta+\sin\alpha ~ p_\gamma \cr
G_3&=-p_\alpha
\end{align}
in which $p_\alpha$, $p_\beta$, and $p_\gamma$ are the conjugate momenta of the
pure gauge variables $\alpha$, $\beta$, and $\gamma$.
By the constraints (\ref{6}), using the explicit forms (\ref{10}), we have to set:
\begin{align}\label{11}
p_\alpha=p_\beta=p_\gamma\equiv 0.
\end{align}
By imposing the constraints, setting $l_s=1$ the Hamiltonian takes the form \cite{kabat,kares}
\begin{align}\label{12}
H=\frac{1}{2\mu} \left(p_r^2+\frac{p_\theta^2}{r^2}+\frac{p_\phi^2}{r^2\cos^2(2\theta)}\right)+\frac{\mu}{8}r^4\sin^2(2\theta)
\end{align}
in which $\mu=m_0/2$, as the reduced mass appearing in the relative motion
of two 0-branes. It is easy to check that the canonical momentum of
$\phi$, $p_\phi$, is conserved. So as expected, the two dimensional angular
momentum is a constant of motion. The equations of motion by (\ref{12}) are
\begin{align}\label{13}
&\mu (\ddot{r} - r \dot{\theta}^2) - \frac{p_\phi^2}{\mu r^3 \cos^2(2\theta)}
+\frac{\mu}{2}r^3 \sin^2(2\theta)=0 \cr
&\mu(r \ddot{\theta} + 2 \dot{r}\dot{\theta} )+\frac{2p_\phi^2 \sin(2\theta)}{\mu r^3 \cos^3(2\theta)} +\frac{\mu}{2}r^3 \sin(2\theta)\cos(2\theta)=0\cr
&\dot{\phi}= \frac{p_\phi}{\mu r^2 \cos^2(2\theta)} .
\end{align}
It is easy to check that  $\theta(t)\equiv 0$, by which the potential is set to zero, the
equations for $(r,\phi)$ would come to the form of a free particle in polar coordinate.
As an illustration that the above equations can develop bound-states,
the plots of a numerical solution are presented in Fig.~1.
In the figure, the outer curve is $r(\phi)$ as the path of the relative motion
of 0-branes in the polar coordinate setup $(r,\phi)$, while the inner curve is a ten
times scaled of $\theta(\phi)$, as the internal degree of freedom causing the
effective attractive force between 0-branes. Evidently, this solution represents an almost
circular path for the relative motion of 0-branes.

\begin{figure}[t]
\begin{center}
\includegraphics[width=0.5\columnwidth]{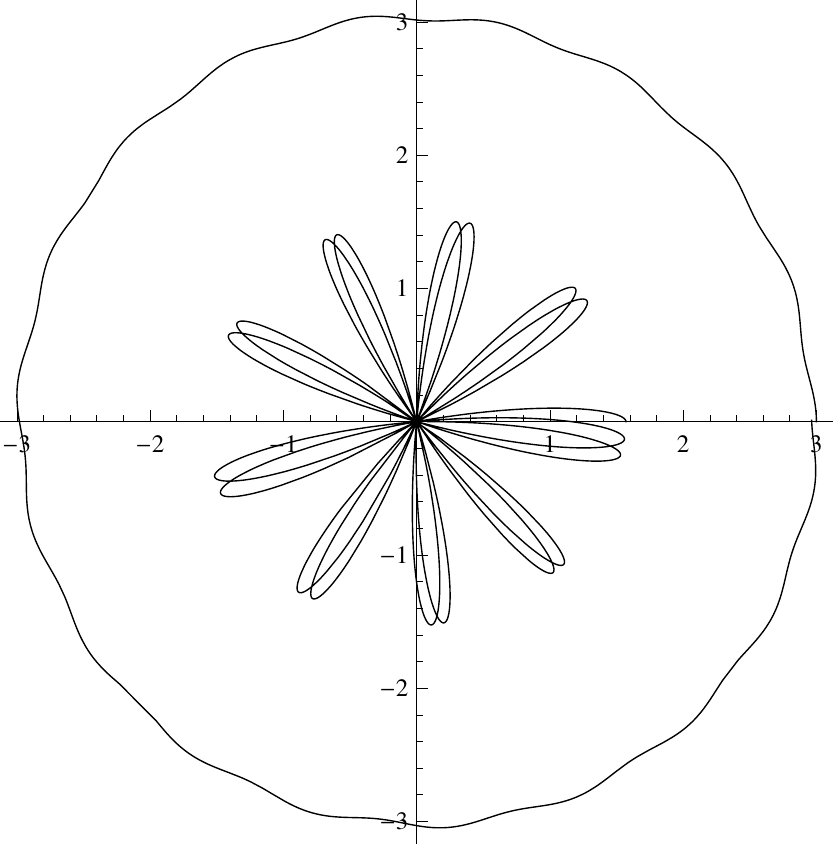}
\caption{\small The plots of a numerical solution of (\ref{13}). The outer curve
is representing the radial coordinate as a function of the polar angle $\phi$.
The inner one, which is scaled ten times to make it visible, is $\theta(\phi)$.
The solution is by the conditions: $\mu=1/2$, $l_s=1$, $p_\phi=1.42$,
$r(0)=3$, $\theta(0)=0.157~$rad, $\dot{r}(0)=\dot{\theta}(0)=0$, $\phi(0)=0$ }
\end{center}
\end{figure}

\section{Quantum Dynamics}
In passing to quantum theory, the constraints in operator form define the
physically acceptable states as
\bea\label{14}
\hat{G}\,|\psi\rangle =0
\eea
By the replacements
\begin{align}\label{15}
p_\alpha\to -\mathrm{i}\,\frac{\partial}{\partial\alpha},~~~~
p_\beta\to -\mathrm{i}\,\frac{\partial}{\partial\beta},~~~~
p_\gamma\to -\mathrm{i}\,\frac{\partial}{\partial\gamma}
\end{align}
one would find, as expected, that the physical wave-functions do not depend on the pure gauge 
degrees of freedom $\alpha$, $\beta$, and $\gamma$. The Laplacian operator can be
constructed using the metric $g_{ij}$
\begin{align}\label{16}
\nabla^2\equiv \frac{1}{\sqrt{g}} \partial_i(\sqrt{g}g^{ij}\,\partial_j)
\end{align}
in which $g=\det g$, explicitly found to be $\frac{1}{4}r^5\sin\gamma\sin(4\theta)$ \cite{kares}.
So, in the coordinate setup $(r,4\,\theta,\phi)$, with $0\leq \theta \leq \pi/4$
and $0\leq \phi \leq 2\pi$, the Hamiltonian acting on the wave-function
$\psi(r,\theta,\phi)$, takes the form \cite{kabat,kares}
\begin{align}\label{17}
H=-\frac{1}{2\mu}\left(\frac{1}{r^5}\partial_r \left(r^5\partial_r\right)+\frac{1}{r^2}\nabla^2_\Omega\right)+\frac{\mu}{8}r^4\sin^2(2\theta),
\end{align}
in which
\begin{align}\label{18}
\nabla^2_\Omega=\frac{1}{\sin(4\theta)}\partial_\theta \left(\sin(4\theta) \partial_\theta\right)+\frac{\partial^2_\phi}{\cos^2(2\theta)}.
\end{align}
Using the scaling $\psi\to r^{-3/2}\psi$, the Hamiltonian comes to the form
\begin{align}\label{19}
H=-\frac{1}{2\mu}\left(\frac{1}{r^2}\partial_r \left(r^2\partial_r\right)+\frac{1}{r^2}(\nabla^2_\Omega-\sfrac{15}{4})\right)+\frac{\mu}{8}r^4\sin^2(2\theta),
\end{align}
By the introduced separation of variables, the two dimensional angular momentum
is $L_z=-\mathrm{i}\,\frac{\partial}{\partial\phi}$ \cite{kares},
and obviously commutes with the Hamiltonian, $[\hat{L}_z,\hat{H}]=0$.
So one can construct states with given energy and angular momentum.

\subsection{Angular momentum spectrum}
Here the aim is to find the eigenfunctions and eigenvalues of the
operator $\nabla^2_\Omega$
\begin{align}\label{20}
\nabla^2_\Omega\; \mathcal{Y}_\lambda(\theta,\phi) = \lambda\; \mathcal{Y}_\lambda(\theta,\phi)
\end{align}
for which we assume as usual
\begin{align}\label{21}
\mathcal{Y}_\lambda (\theta,\phi)=g_\lambda(\theta)\frac{e^{i m_z\phi}}{\sqrt{2\pi}}
\end{align}
with $m_z$ as the quantum number associated to the angular momentum in the
two dimensional configuration space. Although the pure gauge degrees of freedom
have been separated out, it is known that a remaining discrete gauge transformation
would cause that only even integer values are accepted for $m_z$ \cite{kares}.
In particular, the shifts $\alpha\to\pi+\alpha$ and $\phi\to 2\pi+\phi$ would make equal
changes to the original variables, namely $x_{i\,a}\to -x_{i\,a}$, if $m_z$ has an odd value.
So, to construct absolute gauge invariant physical states, the quantum number
$m_z$ has to be even, setting
\begin{align}\label{22}
m_z=2\,m,~~~~~~m=0,\pm 1,\pm2,\cdots.
\end{align}
Using the change of variable $x=\cos(4\theta)$, one has
\begin{align}\label{23}
\frac{\mathrm{d}}{\mathrm{d}x}\left( (1-x^2)
\frac{\mathrm{d}g_\lambda}{\mathrm{d}x}\right)-\frac{m^2}
{2(1+x)}g_\lambda(x)=\frac{\lambda}{16}\;g_\lambda (x).
\end{align}
As the spectrum is invariant under the change $m\to -m$, from now on we take $m\geq 0$.
Using the replacement $g_\lambda(x)=(1+x)^{m/2}Q_\lambda(x)$,
\begin{align}\label{24}
(1-x^2)Q''(x)+\big(m-(m+2)x\big)Q'(x)-\left(\lambda+\frac{m(m+2)}{4}\right)Q(x)=0,
\end{align}
which is known to have Jacobi polynomials of order $n=l-m\geq 0$,
$\mathcal{P}_{n}^{(0,m)}(x)$, as solutions \cite{ryzhik}. By this the
eigenvalue $\lambda$ is found
\begin{align}\label{25}
\lambda=-16(l-m/2)(l-m/2+1),~~~~~~m\leq l=0,1,\cdots,
\end{align}
for the normalized eigenfunction
\begin{align}\label{26}
\mathcal{Y}_l^m(\theta,\phi)=\sqrt{\frac{2l-m+1}{2^{m+1}}}(1+\cos(4\theta))^{m/2}
\mathcal{P}_{l-m}^{(0,m)}(\cos(4\theta))\frac{e^{2 i m\phi}}{\sqrt{2\pi}}
\end{align}
The Jacobi polynomials of our interest satisfy the following recurrence relation, which comes
mostly helpful when the matrix elements of the Hamiltonian (\ref{19}) are evaluated in the angular momentum basis:
\begin{align}\label{27}
\frac{2(l+1)(l-m+1)}{(2l-m+1)(2l-m+2)}\mathcal{P}_{l-m+1}^{(0,m)}(x)+
\frac{2l(l-m)}{(2l-m)(2l-m+1)}\mathcal{P}_{l-m-1}^{(0,m)}(x)\cr
+\frac{m^2}{(2l-m)(2l-m+2)}\mathcal{P}_{l-m}^{(0,m)}(x)=
x\,\mathcal{P}_{l-m}^{(0,m)}(x)
\end{align}

\subsection{Harmonic oscillator solution}
As we are going to evaluate the spectrum of the Hamiltonian (\ref{19}) by
the variational Rayleigh-Ritz method \cite{merzbacher}, a set of basis functions
is needed, for which we shall take those of harmonic oscillator. For a harmonic
oscillator with kinetic term as in (\ref{19}) and unit frequency ($\omega=1$), taking
\begin{align}\label{28}
\psi_{E,l,m}(r,\theta,\phi)=R_{E,l,m}(r)\;\mathcal{Y}_l^m(\theta,\phi)
\end{align}
the radial equation would come to the form
\begin{align}\label{29}
-\frac{1}{2\mu} \left(R''_{E,l,m}-\frac{J_l^m(J_l^m+1)}{r^2} R_{E,l,m}\right) +
\frac{1}{2}\mu r^2 R_{E,l,m} = E\, R_{E,l,m}
\end{align}
in which
\begin{align}\label{30}
J_l^m=4\,l-2\,m+3/2.
\end{align}
It is known that the above has normalized solutions in terms of the Laguerre polynomials
\begin{align}\label{31}
R_{k,l,m}(r)=\sqrt{\frac{2\,k!\, \mu^{J_l^m+3/2}}{\Gamma(k+J_l^m+3/2)}}\;
r^{J_l^m}e^{-\mu r^2/2}\,L_k^{(J_l^m+1/2)}(\mu r^2)
\end{align}
with ($k=0,1,2,\cdots$):
\begin{align}\label{32}
E_{k,l,m}=2k+J_l^m+3/2=2k+4l-2m+3
\end{align}
To calculate the matrix elements of the Hamiltonian (\ref{19}),
the following recurrence relations for Laguerre polynomia1ls would appear mostly
useful \cite{ryzhik}:
\begin{align}\label{33}
L_k^{(\alpha+1)}(x) - L_{k-1}^{(\alpha+1)}(x) &=L_k^{(\alpha)}(x)\cr
(2k+\alpha+1-x)L_k^{(\alpha)}(x)&=
(k+1)L_{k+1}^{(\alpha)}(x)+(k+\alpha)L_{k-1}^{(\alpha)}(x)\cr
(k+\alpha)L_{k-1}^{(\alpha)}(x)-k L_k^{(\alpha)}(x)&=x\;L_{k-1}^{(\alpha+1)}(x)
\end{align}
The following identity for the integral of Laguerre polynomials is known as well \cite{MF}
\begin{align}\label{34}
&\int_0^\infty z^p\; L_{k'}^{(p-\tau')}(z)L_{k}^{(p-\tau)}(z) \mathrm{d}z
 =(-1)^{k' + k} \;\tau'! \;\tau! \;\times \cr
&\sum_{\sigma=\mathrm{max}
\{\!\!\mathop{~}^{k'-\tau'}_{k-\tau}\!\}}^{\mathrm{min}\{\!\!\mathop{~}^{k'}_{k}\!\}}
\frac{(p + \sigma)!}{\sigma! (k' - \sigma)! (k - \sigma)! (\sigma + \tau'-k')!
(\sigma + \tau - k)!}.
\end{align}
Of course if
$\mathrm{max}\{\!\!\mathop{~}^{k'-\tau'}_{k-\tau}\!\}>\mathrm{min}\{\!\!\mathop{~}^{k'}_{k}\!\}$
the integral is zero.

\section{Rayleigh-Ritz Method and Spectrum}
To find the eigenvalues of the Hamiltonian (\ref{19}) we use the Rayleigh-Ritz variational method,
in which a basis function is needed to approximate the exact eigenfunctions.
Here we take the basis function to be a collection of eigenstates of harmonic oscillator
obtained in previous part. As we are interested to find eigenvalues with given
angular momentum $m_z$, the basis function is taken (recall $m_z=2\,m$)
\begin{align}\label{35}
&\Big\{  \psi_{k,l,m_z/2}(r,\theta,\phi)  \Big\},~~~~~~~~~~~~~\mathrm{with}\cr
&~~l=\frac{m_z}{2},\cdots,\frac{m_z}{2}+n_\mathrm{max},~~~
~~k=0,\cdots,n'_\mathrm{max}
\end{align}
in which $n_\mathrm{max}$ and $n'_\mathrm{max}$ determine the level of truncations. By this choice, the number of
the members of the basis function is equal to $(n_\mathrm{max}+1)(n'_\mathrm{max}+1)$.

Before to proceed, it would be useful to determine how the spectrum depends on
the initial parameters $l_s$ and $g_s$ (recall $m_0=1/(g_sl_s)$, and $\mu=m_0/2$).
By the re-scalings \cite{kabat}
\begin{align}\label{36}
X_i\to g_s^{1/3}\, l_s\, X_i,~~~~P_i\to g_s^{-1/3}\,l_s^{-1}\, P_i
\end{align}
in the Hamiltonian (\ref{5}) one finds that the eigenvalues have the form
$E=c\,g_s^{1/3}\,l_s^{-1}$, with $c$ as dimensionless number (recall we have set
$\hbar=c=1$).

\begin{table}[t]{\scriptsize
\begin{center}
\begin{tabular}{c| ccc c c c  || c| ccc c c c }
$m_z$ &  $E_1$ & $E_2$ & $E_3$ & $E_4$ &  $E_5$ &  $E_6$
& $m_z$ &  $E_1$ & $E_2$ & $E_3$ & $E_4$ &  $E_5$ &  $E_6$   \\
\hline
0 & 2.66& 4.54& 5.95& 7.15& 8.25& 9.09 & 22 &13.5& 14.9& 16.5& 18.3& 20.4& 22.8 \\
2 & 4.13& 5.31& 6.22& 7.16& 8.34& 9.79& 24 &14.4& 15.9& 17.6& 19.5& 21.6& 24.1 \\
4 & 5.39& 6.13& 6.89& 7.91& 9.22& 10.9& 26 & 15.4& 16.9& 18.7& 20.6& 22.9& 25.4\\
6 & 6.44& 6.99& 7.83& 8.95& 10.4& 12.1& 28 & 16.3& 17.9& 19.7& 21.8& 24.1& 26.7\\
8 & 7.33& 7.96& 8.89& 10.1& 11.6& 13.4  & 30 & 17.3& 18.9& 20.8& 22.9& 25.3& 28.0 \\
10&8.18& 8.95& 9.97& 11.3& 12.8& 14.7  & 32 & 18.2& 19.9& 21.9& 24.1& 26.5& 29.2\\
12&9.03& 9.95& 11.1& 12.4& 14.1& 16.1  & 34 &19.2& 21.0& 23.0& 25.2& 27.7& 30.5 \\
14&9.90& 10.9& 12.2& 13.6& 15.4& 17.4  & 36 & 20.2& 22.0& 24.0& 26.4& 28.9& 31.8 \\
16&10.8& 11.9& 13.3& 14.8& 16.6& 18.8  & 38 & 21.1& 23.0& 25.1& 27.5& 30.1& 33.1 \\
18&11.7& 12.9& 14.3& 16.0& 17.9& 20.1  & 40 & 22.1& 24.0& 26.2& 28.6& 31.3& 34.3 \\
20 &12.6& 13.9& 15.4& 17.2& 19.1& 21.4 & 42 & 23.1& 25.0& 27.3& 29.8& 32.5& 35.6
\end{tabular}
\caption{{\small The first six energy eigenvalues for given $m_z$
by the Rayleigh-Ritz method, in units $g_s^{1/3}\,l_s^{-1}$.
For each $m_z$ basis function consists of 80 elements. }}
\end{center}
}\end{table}

In calculation of the matrix elements of the Hamiltonian (\ref{19}) one could avoid
explicit integrations over $r$ and $\theta$ variables, simply by using the recurrence
relations (\ref{27}) and (\ref{33}), and the integral identity (\ref{34}).

For the energy eigenvalues reported in Tab.~1
we have set $n_\mathrm{max}=n'_\mathrm{max}=8$,
making 80 elements for the basis function for each $m_z$.

\begin{figure}[t]
\begin{center}
\includegraphics[width=0.8\columnwidth]{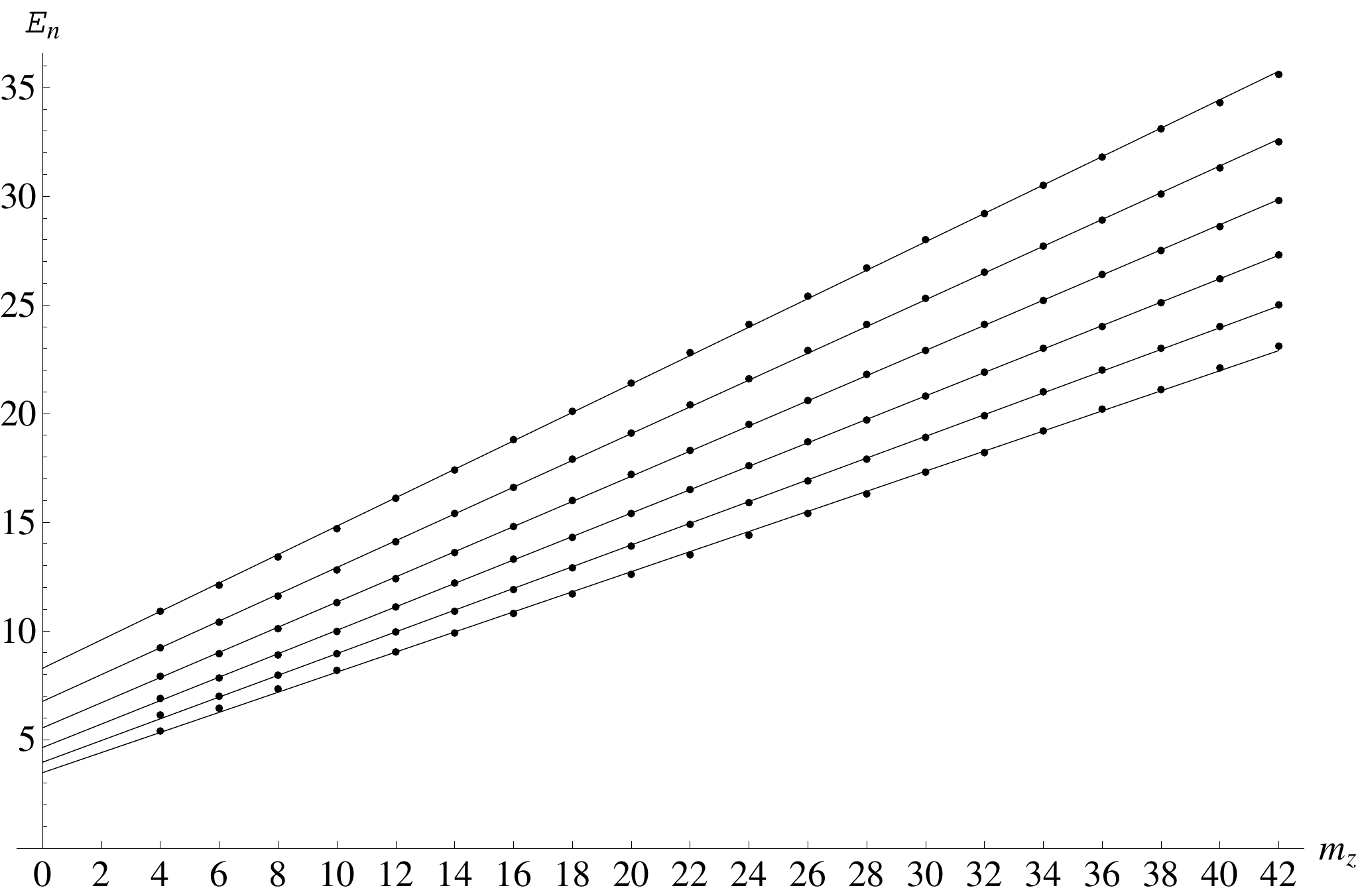}
\caption{\small The plots of energy eigenvalues versus $m_z$, according to
Tab.~1, together with the straight-line fittings given in Eq.~(\ref{37}).}
\end{center}
\end{figure}

Apart from two lowest $m_z$'s, the values given in Tab.~1 together with the straight-line
data fittings are plotted in Fig.~2. The results of the fittings are presented in Eq.~(\ref{37}),
with the brackets indicating the standard error for each given value:
\begin{align}\label{37}
&E_1=3.474 \,{\scriptstyle [0.059]} + 0.462 \,{\scriptstyle [0.002]}\, m_z,~~~
&E_2=3.953 \,{\scriptstyle [0.031]} + 0.500 \,{\scriptstyle [0.001]}\, m_z,\cr
&E_3=4.632 \,{\scriptstyle [0.020]}\, + 0.539 \,{\scriptstyle [0.001]}\, m_z,~~~
&E_4=5.535 \,{\scriptstyle [0.027]}\, + 0.579 \,{\scriptstyle [0.001]}\, m_z,\cr
&E_5=6.754 \,{\scriptstyle [0.038]}\, + 0.616 \,{\scriptstyle [0.001]}\, m_z,~~~
&E_6=8.277 \,{\scriptstyle [0.047]}\, + 0.654 \,{\scriptstyle [0.002]}\, m_z.
\end{align}
By the present standard errors one finds that all the percentage errors are less than $\%2$.
Further, all the statistical P-values for the straight-line fittings are less than $10^{-22}$, leaving
almost no room for the null hypothesis.

\subsection{Light-cone interpretation}
The obtained spectrum may
be interpreted as the one for massive 0-branes in 2+1 dimensions, or
in a Matrix theory perspective \cite{9610043}, as for massless
particles in 3+1 dimensions but in the light-cone frame. For the latter way of interpretation,
the relative motion of bound-state constituents (defined by
$\vec{P}_\perp=\sum_i\vec{k}_{\perp i}=0$) is related to the
masses' constituents and the potential $W$ in transverse directions
as following \cite{lc}
$$
H:=P^-=\sum_{i=1,2} \frac{\vec{k}_{\perp i}^2+{m_i}^2}{2\,p^+_i}+\frac{W}{2\,P^+},
$$
in which $P^+=\sum_i p^+_i$, and $p^+_i$'s appear as the
masses of constituents but in the transverse directions of the light-cone frame.
For the case of interest, setting $m_i=0$ and $p^+_1=p^+_2$ one simply has
the relation between Hamiltonian eigenvalues for relative motion
($\vec{P}_\perp=0$) and the mass squared of bound-state
using the key relation $\mathcal{M}^2=2\,P^- P^+$ \cite{9610043,lc}.
So in the light-cone frame the linear relation between
the Hamiltonian eigenvalues and the angular momentum turns as
the linear relation between the mass squared and the angular momentum, just
reminiscent the observed one in hadron physics.

\vskip 0.5cm
{\bf Acknowledgement: }
This work is supported by the Research Council of Alzahra University.



\end{document}